\documentclass[11pt]{article}
\newcommand{\dis}{\displaystyle}

\newcommand{\tA}{\theta^\alpha}

\newcommand{\xM}{x^\mu}
\newcommand{\vfmn}{\varphi_{\mu\nu}}
\newcommand{\vfma}{\varphi_{\mu\alpha}}
\newcommand{\vfab}{\varphi_{\alpha\beta}}

\newcommand{\Gmab}{(C \Gamma_\mu)_{\alpha\beta}}
\newcommand{\GMab}{(C \Gamma^\mu)_{\alpha\beta}}

\newcommand{\be}{\begin{equation}}
\newcommand{\bae}{\begin{eqnarray}}
\newcommand{\ee}{\end{equation}}
\newcommand{\eae}{\end{eqnarray}}

\begin{document}

\thispagestyle{empty}

\vspace*{-1cm}
\begin{flushright}
FTUV--01-0505 \quad IFIC--01-24
\\
May 5, 2001
\\[1cm]
\end{flushright}

\begin{center}
\begin{Large}
\bfseries{Superalgebra cohomology, the geometry of
  extended superspaces and superbranes}
\end{Large}

%\vspace*{0.6cm}
\vspace*{0.4cm}

\begin{large}
J.~A.~de~Azc\'arraga$^1$ and J.~M. Izquierdo$^2$
\footnote{Invited lecture delivered at the XXXVII Karpacz
  Winter School on {\it New Developments in Fundamental Interaction Theories},
  February 6-15, 2001, Karpacz, Poland. To be published in Proceedings.\\
e-mail addresses:
j.a.de.azcarraga@ific.uv.es, izquierd@fta.uva.es}
\end{large}

\vspace*{0.6cm}
\begin{it}
$1$ Departamento de F\'{\i}sica Te\'orica, Universidad de Valencia
\\
and IFIC, Centro Mixto Universidad de Valencia--CSIC,
\\
E--46100 Burjassot (Valencia), Spain
\\[0.4cm]
$2$ Departamento de F\'{\i}sica Te\'orica, Universidad de Valladolid
\\
E--47011 Valladolid, Spain\\
\end{it}

\end{center}
\vspace*{1cm}
\begin{abstract}
We present here a cohomological analysis of the new spacetime superalgebras
that arise in the context of superbrane theory. They lead to enlarged
superspaces that allow us to write D-brane actions in terms of fields 
associated with the additional superspace variables. This suggests that 
there is an extended superspace/worldvolume fields democracy for 
superbranes.
\end{abstract}

\setcounter{footnote}{0}
\renewcommand{\thefootnote}{\arabic{footnote}}

\newpage

%%%%%%%%%%%%%%%%%%%%%%%%%%%%%%%%%%%%%%%%%%%%%%%%%%%%%%%%%%%%%%%%%
%%%%%%%%%%%%%%%%%%%%%%%%%%%%%%%%%%%%%%%%%%%%%%%%%%%%%%%%%%%%%%%%%
%%%%%%%%%%%%%%%%%%%%%%%%%%%%%%%%%%%%%%%%%%%%%%%%%%%%%%%%%%%%%%%%%%%%%%%%%
\section{Introduction}

Due to the development of the new string theory, it has become clear that
the supersymmetry algebra contains new bosonic tensorial generators, which 
are central if one does not consider the Lorentz part of the complete
algebra. The study of the structure of the supersymmetry algebra goes back to
\cite{azca-HLS75}, and tensorial charges were already considered in 
\cite{azca-HP82} 
(see also \cite{azca-DF82, azca-Z84}). An enlarged 
superalgebra with an additional `central'
fermionic generator was introduced in \cite{azca-G89}; other, more general
algebras were considered in \cite{azca-BS95}, where it was proved that
to every super-$p$-brane of the branescan in \cite{azca-AETW87}, corresponds
a new spacetime superalgebra, generalizing the results of 
\cite{azca-G89, azca-S94}. The point of view in \cite{azca-BS95,azca-S97} 
constitutes the Lie superalgebra counterpart of the Chevalley-Eilenberg (CE) 
supersymmetry algebra cohomology (see \cite{azca-AI95,azca-AI01}) 
analysis of the scalar branes previously done in \cite{azca-AT89}. 
We report here on a recent work \cite{azca-CAIP00} 
which leads to a systematic cohomological construction of 
all these algebras, with both bosonic and fermionic generators, as
well as their associated extended superspace $\tilde \Sigma$ {\it groups}.
These contain additional coordinates, besides those ($x^\mu,\theta^\alpha$)
of the standard superspace $\Sigma$, which can be used to construct 
manifestly invariant super-$p$-brane Wess-Zumino (WZ) terms
\cite{azca-S94,azca-BS95,azca-S97,azca-CAIP00}. Indeed,
it is well known (see {\it e.g.} \cite{azca-AI95}) that the 
quasi-invariance of Lagrangians (invariance but for a total
derivative) exhibits the non-trivial cohomology of the symmetry 
group, and that they can be rendered manifestly invariant by using 
the additional variables associated with the extended group. 
This is reflected in the realization of the
symmetries in terms of Noether currents and charges, and we shall 
provide their general expression for superbranes. As one 
might expect, the new variables on $\tilde\Sigma$ appear trivially 
(in total derivatives) in the action of the scalar super-$p$-branes. 
In the case of D-branes, however, some new variables appear 
non-trivially since the worldvolume fields can be constructed as 
pull-backs of suitably enlarged superspaces that correspond to 
new superalgebras with additional fermionic generators.

\section{Extended superspaces given by central extensions of the 
supertranslation group}
\label{azca-CESA}
%%%%%%%%%%%%%%%%%%%%%%%%%%%%%%%%%%%%%%%%%%%%%%%%%%%%%%%%%%%%%%%%%

Standard superspace itself provides the simplest example of our point of
view. Consider the  abelian, odd, {\it supertranslation group} 
$\textrm{sTr}_D$, of group law  
$\theta''^\alpha = \theta'^\alpha + \theta^\alpha$ (generically, 
we denote a group law as $g''=g'g\equiv L_{g'}g=R_g g'$; $L$ and
$R$ are the left and right actions of the group on itself).
Associated with $\textrm{sTr}_D$ is the trivial Maurer-Cartan (MC) 
equation $d \Pi^\alpha = 0\,$ ($\Pi^\alpha=d\theta^\alpha$), 
which is the dual version of the 
corresponding abelian Lie superalgebra $\{D_\alpha,D_\beta\}$=0
\footnote{We use $D_\alpha$ (covariant derivatives) rather than 
$Q_\alpha$ (supersymmetry generators) because we deal with LI
(hence, supersymmetry invariant) forms and vector fields, but this 
is unessential: the left and right algebras have the same structure 
constants but for an overall sign that may be conveniently ignored 
here.}. Now, let $\theta^\alpha$ be Majorana. Then,
$(C \Gamma^\mu)_{\alpha\beta} \Pi^\alpha\wedge \Pi^\beta$ defines a 
{\it $\textrm{Tr}_D$-valued non-trivial CE two-cocycle} since
a) it is closed and left-invariant (LI) {\it i.e.}, it is a CE
cocycle and b) it is not $d$ of a LI form (not a coboundary).
Therefore, it is consistent to extend 
$d\Pi^\alpha=0$ by a one-form ${\tilde\Pi}^\mu$ 
(${\tilde\Pi}^\mu\equiv (1/2)
(C\Gamma^\mu)_{\alpha\beta} \theta^\alpha d\theta^\beta$, say) so that
\begin{equation}
d {\tilde\Pi}^\mu = (1/2) (C \Gamma^\mu)_{\alpha\beta}
            \Pi^\alpha \Pi^\beta \quad .
\label{azca-2.3}
\end{equation}
Clearly, ${\tilde \Pi}^\mu$ and $\Pi^\alpha$ define a free differential 
algebra (FDA) but they are not the MC one-forms of a {\it Lie} algebra since
{\it ${\tilde \Pi}^\mu$ is not LI}. To remedy this, we introduce
a {\it new group coordinate $x^\mu$ -- Minkowski space $\textrm{Tr}_D$--} 
and define instead \begin{equation}
\Pi^\mu =dx^\mu + {\tilde\Pi}^\mu= 
dx^\mu + (1/2) (C \Gamma^\mu)_{\alpha \beta}
          \theta^\alpha d\theta^\beta \quad ,
\quad \mu=0,1,\dots , D-1 \quad .
\label{azca-2.3b}
\end{equation}
Obviously, $d\Pi^\mu=d{\tilde \Pi}^\mu$ and we may now
choose the transformation law for 
$x^\mu$ so that $\Pi^\mu$ is LI,
\begin{equation}
x''^\mu = x'^\mu +x^\mu -(1/2) (C \Gamma^\mu)_{\alpha\beta}
            \theta'^\alpha \theta^\beta
\quad.
\label{azca-xgcl}
\end{equation}
This gives \cite{azca-AA85,azca-AI95} {\it rigid superspace} 
$\Sigma$ as a {\it group} parametrized by 
$(\theta^\alpha,x^\mu)$, with group law now given by 
$\theta''^\alpha = \theta'^\alpha + \theta^\alpha$ and (\ref{azca-xgcl}): 
{\it supersymmetry is the result of the non-trivial cohomology of the
odd supertranslation group}.

The philosophy behind this simple superspace example, that 
`fermions ($\theta$'s) are first' and that {\it rigid superspaces are 
group extensions}, may be extended by considering other types
of two-cocycles ({\it i.e}, valued on more general spaces than
$Tr_D$) on the $sTr_D$ algebra. Explicitly,  given a particular 
{\it sTr}${}_D$ algebra to be extended, one

a)
looks for a non-trivial CE two-cocycle of the desired 
Lorentz-covariant nature. This means searching for Lorentz-tensor-valued 
LI closed two-forms that are not $d$(LI one-form).

b)
Introduces a new LI one-form, the differential of which is
the two-cocycle. Then,

c)
the left invariance of the new one-form is achieved by fixing the 
transformation properties of the {\it new group coordinate}. This defines
in general an \emph{extended superspace (super)group} manifold 
${\tilde \Sigma}$.

d)
The new LI one-form together with the 
MC equations  
automatically define by (LI one-forms/LI vector fields)
duality an extended Lie algebra.

e)
Since the required Lorentz group symmetry is implicit in the process, 
the extension cocycles must be covariant under the action 
of $Spin(1,D-1)$.

The extension procedure described above can be applied more than
once. 
\medskip

Consider the {\it general case of `central' bosonic extensions} of 
$\textrm{sTr}_D$ (they are 
really tensorial, since the Lorentz generators do not commute with the
`central' ones). As in the above superspace example, we may look at the
problem from the Lie algebra $\cal G$ or the group $G$ point of view:

\medskip
\noindent 1) {\it Lie algebra extension point of view}

\noindent
When described in terms of LI forms, the algebra extensions require the 
existence of higher order ($\alpha,\beta$)-{\it symmetric} Lorentz tensors 
$(C\Gamma^{\mu_1\dots\mu_p})_{\alpha\beta}$ of rank $p$
\begin{equation}
d\Pi^{\mu_1\dots \mu_p} \equiv (1/2)
(C \Gamma^{\mu_1\dots \mu_p})_{\alpha\beta} \Pi^\alpha \Pi^\beta
\label{azca-2.4}
\end{equation}
$(\Pi^\alpha\Pi^\beta\equiv\Pi^\alpha\wedge\Pi^\beta
=\Pi^\beta\wedge\Pi^\alpha$; we omit wedge products). The corresponding 
generators $Z_{\mu_1 \dots \mu_p}$ are all $(D_\alpha-$){\it central}, as 
the translation generator $X_{\mu}=P_\mu$ itself,
and are associated with new central charges.

The LI of the new forms in~(\ref{azca-2.4})
\emph{requires} new group parameters
$\varphi^{\mu_1 \dots \mu_p}$ so that
\begin{equation}
\Pi^{\mu_1\dots \mu_p} =
d \varphi^{\mu_1 \dots \mu_p}  +
(1/2)(C \Gamma^{\mu_1\dots \mu_p})_{\alpha \beta}
\theta^\alpha \Pi^\beta
\label{azca-2.8}
\end{equation}
is LI. These new parameters $\varphi^{\mu_1\dots \mu_p}$ generalize the 
spacetime parameters $x^\mu$, and their associated 
generators $Z_{\mu_1\dots\mu_p}$
may be considered as {\it generalised momenta}.
There are no (two-cocycle) restrictions coming from the 
Jacobi identity since the r.h.s of (\ref{azca-2.4}) is trivially consistent 
with $d(d\Pi^{\mu_1\dots \mu_p})\equiv 0$.

\medskip
\noindent 2) {\it Group extension point of view}

The closedness of the r.h.s. of (\ref{azca-2.4}) means that the Lorentz 
tensor-valued two-cocycle on
sTr${}_D$, $\xi^{\mu_1\dots \mu_p} (\theta',\theta)=
\theta'^\alpha (C \Gamma^{\mu_1\dots \mu_p})_{\alpha \beta} 
\theta^\beta$,
satisfies also (trivially) the {\it two-cocycle condition}
\begin{equation}
\xi(\theta,\theta')^{\mu_1\dots \mu_p} +
\xi(\theta + \theta' ,\theta'')^{\mu_1\dots \mu_p} =
\xi(\theta , \theta' + \theta'')^{\mu_1\dots \mu_p} +
\xi(\theta' ,\theta'')^{\mu_1\dots \mu_p}
\label{azca-cocycle}
\end{equation}
The symmetry of $(C\Gamma ^{\mu_1 \dots \mu_p})_{\alpha \beta}$ 
is needed to prevent the above two-cocycle from being trivial,
since the possible function $\eta^{\mu_1\dots \mu_p}(\theta)$
on sTr${}_D$ that might generate the 
two-coboundary ($\xi_{cob}^{\mu_1\dots\mu_p}(\theta{'},\theta)=
\eta^{\mu_1\dots\mu_p}(\theta+\theta{'})-\eta^{\mu_1\dots\mu_p}(\theta{'})
-\eta^{\mu_1\dots\mu_p}(\theta)$) is zero: 
$\eta^{\mu_1\dots\mu_p}(\theta)=
\theta^\alpha(C\Gamma^{\mu_1\dots\mu_p})_{\alpha\beta}\theta^\beta
\equiv 0$. Hence, 
{\it The problem of finding all \emph{central}
extensions of the $sTr_D$ algebra 
$\{D_\alpha,D_\beta\}=0$ 
is reduced to finding a basis of the symmetric space 
$\Pi^{(\alpha}\otimes \Pi^{\beta)}$ in terms of $p$-Lorentz tensors 
$(C \Gamma^{\mu_1\dots \mu_p})_{\alpha\beta}$ symmetric in} 
($\alpha,\beta$).

The answer for different spacetime dimensions $D$ depends on the properties
of their respective $\Gamma$ matrices, since they determine the existence
of non-trivial cocycles (see \cite{azca-CAIP00} for a table). We shall only
consider here the example of the

%%%%%%%%%%%%%%%%%%%%%%%%%%%%%%%%%%%%%%%%%%%%%%%%%%%%%%%%%%%%%%%%%%
%%%%%%%%%%%%%%%%%%%%%%%%%%%%%%%%%%%%%%%%%%%%%%%%%%%%%%%%%%%%%%%%%%%%%%

\subsection{ D=11, M-theory extended superspace}

The maximally centrally extended FDA is obtained by adding $\Pi^\mu$, 
$\Pi^{\mu_1\mu_2}$, $\Pi^{\mu_1\dots\mu_5}$ to $\Pi^\alpha=d\theta^\alpha$
($\theta^\alpha$ Majorana, $\alpha=1,\dots,32$) satisfying
\begin{equation}
\begin{array}{l}
\displaystyle
d\Pi^\alpha =0 \; , \;
d\Pi^\mu = {1\over 2} (C \Gamma^\mu)_{\alpha \beta}
\Pi^\alpha \Pi^\beta \quad,
\\
\displaystyle
d\Pi^{\mu_1\mu_2} =
{1\over 2} (C \Gamma^{\mu_1\mu_2})_{\alpha \beta} \Pi^\alpha \Pi^\beta
\; ,\;
d\Pi^{\mu_1 \dots \mu_5} =
{1\over 2} (C \Gamma^{\mu_1 \dots \mu_5})_{\alpha \beta}
\Pi^\alpha \Pi^\beta
\quad.
\end{array}
\label{azca-2.10}
\end{equation}
There are no 
{\it one}-forms $\Pi^{\mu_1\mu_2}$, $\Pi^{\mu_1 \dots \mu_5}$
LI on $\Sigma$. They
can be made LI by introducing {\it new} `central' (tensorial) coordinates
$\varphi^{\mu_1\mu_2}$, $\varphi^{\mu_1 \dots \mu_5}$. These \emph{define}
the $D=11$, $N=1$ extended superspace
$\tilde\Sigma(\theta^\alpha, x^\mu,\varphi^{\mu_1\mu_2}, \varphi^{\mu_1 \dots
\mu_5})$.
In terms of the central generators
 $X_\mu =
\partial/\partial x^\mu$, $Z_{\mu_1\mu_2}$=
$\partial/\partial \varphi^{\mu_1 \mu_2}$,
$Z_{\mu_1 \dots \mu_5}$=
$\partial/\partial \varphi^{\mu_1 \dots \mu_5}$,
the {\it $D=11$ supersymmetry M-algebra} 
dual to (\ref{azca-2.10}) is 
\begin{equation}
\{D_\alpha,D_\beta\} = (C \Gamma^{\mu})_{\alpha\beta} X_{\mu}
+(C \Gamma^{\mu_1\mu_2})_{\alpha\beta} Z_{\mu_1\mu_2}
+ (C \Gamma^{\mu_1 \dots \mu_5})_{\alpha\beta} Z_{\mu_1 \dots \mu_5}
\quad.
\label{azca-2.11a}
\end{equation}
This is usually referred to \cite{azca-T96}
as the \emph{$M$-theory superalgebra}.
The group law of the extended superspace 
$\tilde \Sigma(\theta^\alpha, x^\mu,\varphi^{\mu_1\mu_2}, \varphi^{\mu_1 \dots
\mu_5})$ is obtained easily as the simplest $\Sigma$ case 
({\it cf}. (\ref{azca-2.3b})).

For a recent discussion of the M-algebra, in which the tensorial
central charges are considered as bilinears of spinors, 
see \cite{azca-BAIL01}.

%%%%%%%%%%%%%%%%%%%%%%%%%%%%%%%%%%%%%%%%%%%%%%%%%%%%%%%%%%%%%%%%%
%%%%%%%%%%%%%%%%%%%%%%%%%%%%%%%%%%%%%%%%%%%%%%%%%%%%%%%%%%%%%%%%%
\section{Non-central additional generators and their extended superspaces}
\label{azca-NCESA}
%%%%%%%%%%%%%%%%%%%%%%%%%%%%%%%%%%%%%%%%%%%%%%%%%%%%%%%%%%%%%%%%%
The above are {\it central} extensions of the basic odd abelian algebra 
$\{ D_\alpha,D_\beta\}=0$ by {\it bosonic} tensorial generators.
But there are also extensions by fermionic generators that make non-abelian
{\it e.g.}, the $[X_\mu,D_\alpha]=0$ commutator. The CE cohomology analysis is
also useful here.
Let us start from a centrally extended superspace 
$\tilde\Sigma(\theta^\alpha,x^\mu,\varphi_{\mu_1\dots\mu_p})$, 
$p$ fixed, and LI one-forms $\Pi^\mu$, $\Pi^\alpha$,
$\Pi_{\mu_1\dots \mu_p}$, satisfying the MC eqs.
\begin{equation}
d \Pi^\mu = a_s (C \Gamma^\mu)_{\alpha\beta}
            \Pi^\alpha \Pi^\beta
\quad,\quad
d\Pi_{\mu_1\dots \mu_p} \equiv a_0
(C \Gamma_{\mu_1\dots \mu_p})_{\alpha\beta} \Pi^\alpha \Pi^\beta
\quad ,
\label{azca-dPimu}
\end{equation}
where $a_s$, $a_0$ are not fixed for convenience.
A non-trivial CE two-cocycle 
with $p$ indices has to be of the type 
$(\mu_1 \dots \mu_{p-1}\alpha_1)$ and, hence, the only available LI 
{\it two}-forms (in this case, fermionic) are 
\begin{equation}
\rho^{(1)}_{\mu_1 \dots \mu_{p-1} \alpha_1}=(C \Gamma_{\nu \mu_1 \dots
\mu_{p-1}})_{\beta \alpha_1} \Pi^\nu \Pi^\beta
\ , \ 
\rho^{(2)}_{\mu_1 \dots \mu_{p-1} \alpha_1}= (C \Gamma^\nu)_{\beta
\alpha_1} \Pi_{\nu \mu_1 \dots \mu_{p-1}} \Pi^\beta
\label{azca-cand1}
\quad .
\end{equation}
For $p=1$, both are closed. For $p \geq 2$, the condition
$d (\rho^{(1)} + \lambda_2 \rho^{(2)})=0$ fixes
$\displaystyle \lambda_2=a_s/ a_0$ 
provided
\begin{equation}
(C \Gamma^\nu)_{(\alpha \beta} (C \Gamma_{\nu \mu_1 \dots
\mu_{p-1}})_{\gamma \delta)}=0
\quad,
\label{azca-Gammaprop}
\end{equation}
which holds for the $(D,p)$ of the scalar branescan 
\cite{azca-AETW87}. Condition (\ref{azca-Gammaprop})
is a {\it new} feature of the {\it `non-central' case}; in the central
(bosonic two-cocycles) case, 
the closedness was {\it trivially} satisfied and hence there was {\it no
condition on $D$; only $p$} (the rank $p$ of the Lorentz tensor,
see below (\ref{azca-2.8})) was restricted by the $(\alpha\beta)$ 
symmetry of the tensor. We
may introduce now a new one--form 
$\Pi_{\mu_1 \dots \mu_{p-1} \alpha_1}$ with
\begin{equation}
d\Pi_{\mu_1 \dots \mu_{p-1} \alpha_1} = a_1 \left((C \Gamma_{\nu
\mu_1 \dots \mu_{p-1}})_{\beta \alpha_1} \Pi^\nu \Pi^\beta\right.
+ {a_s \over a_0} \left. (C\Gamma^\nu)_{\beta \alpha_1} \Pi_{\nu \mu_1
\dots \mu_{p-1}} \Pi^\beta)\right)
\label{azca-dPim1a1}
\end{equation}
(for $p=1$ the coefficient of the second term can be arbitrary).
 This MC equation implies that both $\left[D_\alpha,X_\mu\right]$
and $\left[D_\alpha,Z^{\mu_1 \dots \mu_p}\right]$  are 
modified by a
term proportional to $Z^{\mu_1 \dots \mu_{p-1} \alpha_1}$, the
latter being the only central generator at this stage ($Z^{\mu_1
\dots \mu_{p-1} \alpha_1}$ is central because,
by construction, $\Pi_{\mu_1 \dots
\mu_{p-1} \alpha_1}$ cannot appear at the r.h.s. of a MC equation
expressing the differential of a LI form).

The general features of the extensions with non-central fermionic 
generators are: 

a) The extension two-cocycles (two-forms) may be {\it fermionic} (eqs. 
(\ref{azca-cand1})). This leads to non-zero [bosonic,fermionic] commutators.

b) At any stage in the chain of extensions, the only {\it central}
generator present is the one introduced in the {\it last} extension.

c) Successive central extensions substitute one spinorial index for a vectorial 
one. This leads to one-forms of the type
\begin{equation}
      \Pi_{\mu_1\dots\mu_{p-k}\alpha_1\dots\alpha_k}\equiv\Pi_{\rho_k}
      \ ,\quad \rho_k\equiv(\mu_1\dots \mu_{p-k}\alpha_1\dots \alpha_k)\quad ,
                                    \label{azca-ncn}
\end{equation}
where $\rho_k$ labels the additional coordinates of the 
extended superspace $\tilde \Sigma$.

d) The procedure ends when the $p$ vector indices have become spinorial ones
so that $\Pi_{\rho_k}\rightarrow\Pi_{\rho_p}\equiv\Pi_{\alpha_1\dots\alpha_p}$.

e) For a given $p$, there are consistency conditions that restrict the
spacetime dimension D; for instance, the Green algebra exists for
D=$3$,$4$,$6$ and $10$ only \cite{azca-G89}.

All the extended superspaces have a natural fibre bundle structure
that is inherited from their group extension character; we refer
to  \cite{azca-CAIP00} (see also \cite{azca-AA85}) for details.

%%%%%%%%%%%%%%%%%%%%%%%%%%%%%%%%%%%%%%%%%%%%%%%%%%%%%%%%%%%%%%
\subsection{Two applications: the GS superstring
and the supermembrane}
\label{azca-p12}
%%%%%%%%%%%%%%%%%%%%%%%%%%%%%%%%%%%%%%%%%%%%%%%%%%%%%%%%%%%%%%%%%

Consider the Green-Schwarz superstring case ($p$=1, D=10, N=1).
We shall denote by $\varphi_\mu$ the additional vector parameter and by
$Z^\mu$, $\Pi^{(\varphi)}_\mu$ the associated generator and LI form 
\cite{azca-CAIP00}. The MC eqs. are 
\begin{eqnarray}
d\Pi^\alpha & = & 0 
\ ,\quad 
d \Pi^\mu   =   (1/2) (C \Gamma^\mu)_{\alpha\beta} \Pi^\alpha
\Pi^\beta \;,
\nonumber
\\
d \Pi_\mu^{(\varphi)} & = & (1/2) (C \Gamma_\mu)_{\alpha\beta}
\Pi^\alpha \Pi^\beta
\ ,\;
d \Pi_\alpha  =   (C \Gamma_\mu)_{\alpha\beta}
\Pi^\mu \Pi^\beta
+  (C \Gamma^\mu)_{\alpha\beta}\Pi_\mu^{(\varphi)} \Pi^\beta
\, ;
\label{azca-ssFDA}
\end{eqnarray}
$\mu=0, \dots , 9$, and all spinors here
are MW ($\theta^\alpha\equiv\mathcal{P}_+\theta^\alpha$, $\Pi^\alpha
\equiv\mathcal{P}_+d\theta^\alpha$; notice that $\Pi^\alpha$ and 
$\Pi_\alpha$ are unrelated). The two terms in the r.h.s.
of the last of~(\ref{azca-ssFDA}) are individually closed
($d(d\Pi_\alpha)=0$ follows from  
(\ref{azca-Gammaprop}) for $p=1$, {\it i.e.} by 
$(C \Gamma^\mu)_{(\alpha\beta} (C \Gamma_\mu)_{\gamma\delta)}=0$) 
and hence their relative normalization cannot be fixed by requiring
$d(d\Pi_\alpha)=0$. 

The corresponding Lie superalgebra contains an additional {\it fermionic}
central generator, $Z^\beta$, and is given by
\begin{eqnarray}
\{ D_\alpha, D_\beta\} &=& \GMab X_\mu + \Gmab Z^\mu
\, , 
\nonumber\\
 \left[D_\alpha, X_\mu\right] &=&  \Gmab Z^\beta
\, , \quad
\left[D_\alpha, Z^\mu\right] = \GMab Z^\beta
  \quad ;
\label{azca-ssLa}
\end{eqnarray}
if one omits $Z^\mu$, it reduces to the {\it Green algebra} 
\cite{azca-G89}\footnote{This algebra may be viewed as a `stabilising 
deformation' of $\Sigma$ \cite{azca-C01}. In this context, stability
is achieved by exhausting the second Lie algebra cohomology group
(the non-trivial two-cocycle space) {\it i.e.},
by extending maximally under certain conditions, as done in 
\cite{azca-CAIP00}. This means, {\it e.g.}, including {\it both} generators
$Z_{\mu\nu}$ and $Z_{\mu_1\dots\mu_5}$ in (\ref{azca-2.11a}) if only 
bosonic ones are considered ({\it cf}. \cite{azca-S97}), 
and similarly for the other cases.}.
Note that {\it $X_\mu$ is no longer central} due to the presence of $Z^\beta$.
The associated group manifold is the {\it GS superstring extended
superspace} $\tilde \Sigma(\theta^\alpha,
x^\mu,\varphi_\mu,\varphi_\alpha)$; its group law is given and discussed
in \cite{azca-CAIP00}.

As mentioned, the extended superspaces are suitable to define 
manifestly invariant WZ terms. For instance, using the
 LI forms on ${\tilde\Sigma}
(\theta^\alpha,x^\mu,\varphi_\mu,\varphi_\alpha)$, one obtains the
{\it manifestly invariant WZ term for the GS superstring} 
({\it cf}. \cite{azca-S94})
\begin{equation}
S_{WZ} =\int_W \phi^{*}(\tilde{b}) =   
\int_W \phi^{*}(\Pi_\mu^{(\varphi)} 
 \Pi^\mu + {1 \over 2} \Pi_\alpha  \Pi^\alpha) \quad,
\label{azca-ssWZt}
\end{equation}
$d\tilde{b}=db=h= \Gmab \Pi^\mu \Pi^\alpha \Pi^\beta$ and hence 
$\phi^*(\tilde{b})$ and the standard WZ term $\phi^*(b)$  are 
equivalent; $b$ and $\tilde b$ {\it differ only by an exact form}.

Similarly, it is also possible to write a manifestly invariant D=11
membrane ($p$=2) WZ term. It exists on the
{\it $D=11$ supermembrane extended superspace group 
${\tilde\Sigma}(\tA,\xM,\vfmn,\vfma,\vfab)$}, and is found to be
\begin{equation}
\tilde{b}=(2/3) \Pi_{\mu \nu} \Pi^\mu \Pi^\nu 
- (3/5) \Pi_{\mu \alpha} \Pi^\mu \Pi^\alpha
- (2/15) \Pi_{\alpha \beta} \Pi^\alpha \Pi^\beta\ ,
\label{azca-smbt}
\end{equation}
as given in \cite{azca-BS95}. Again,
$\tilde b$ {\it depends on the additional
variables $\varphi$ through total differentials} 
since $d \tilde{b}=db=h
= (C \Gamma_{\mu\nu})_{\alpha\beta} \Pi^\mu \Pi^\nu \Pi^\alpha \Pi^\beta$. 
The non-WZ part of the action does not depend on the additional 
variables of extended superspace, and remains the standard one. As we 
shall see, this situation will change for D-branes.

%%%%%%%%%%%%%%%%%%%%%%%%%%%%%%%%%%%%%%%%%%%%%%%%%%%%%%%%%%%%%%%%%
\section{ New Noether currents and charges}
\label{azca-noecur}

Let us now give the general expression for the Noether
currents for the additional symmetries $j^i_{\sigma_l}$ 
(see (\ref{azca-ncn}) for the notation) using the manifestly
invariant WZ forms ${\tilde {\cal L}}_{WZ}$ 
defined on the various extended superspaces $\tilde\Sigma$. The Lagrangian
${\tilde {\cal L}}_{WZ}(\xi)$ on the worldvolume $W$ (of coordinates
$\xi^i$, $i=0,1,\dots,p$) is the pull-back 
$\phi^*({\tilde {\cal L}}_{WZ})=
{\tilde {\cal L}}_{WZ} (\xi) d^{p +1} \xi$ 
of the ($p$+1)-form ${\tilde {\cal L}}_{WZ}$ on $\tilde\Sigma$ 
by the map $\phi:W\longrightarrow {\tilde\Sigma}$. 
The charges that correspond to the current densities 
$j^i_{\sigma_l}(\xi)$ appear on the r.h.s. of the
supersymmetry algebra.

The WZ part of the action is $\int
{\tilde {\cal L}}_{WZ} (\xi) d^{p +1} \xi $ and, 
{\it since only ${\tilde{\cal L}}_{WZ}$ depends on the additional 
variables of the extended superspace $\tilde\Sigma$}
(different from ($x^\mu$, $\theta^\alpha$) of $\Sigma$),
we shall focus on ${\tilde{\cal L}}_{WZ}$. We find first 
the general expression for the Noether currents
and then apply it to the simple cases of the GS superstring and the 
supermembrane.

We start by writing the manifestly invariant density 
$\tilde{\cal L}_{WZ} (\xi)$ as
\begin{equation}
\tilde{\cal L}_{WZ} (\xi) 
\equiv \Pi_{\rho_k i} (\xi) \, \Lambda^{\rho_k i} (\xi)
\quad ,\quad \xi^i=(\tau,{\bf \sigma})\quad,\quad i=0,1,\dots ,p\quad ,
\quad 
\label{azca-WZinvar}
\end{equation}
(see (\ref{azca-ncn})) where $\Lambda^{\rho_k}$ is defined by
(\ref{azca-WZinvar}) and denotes the LI $(p$-)form
\begin{equation}
\Lambda^{\rho_k }\equiv
\Lambda^{\mu_1\dots\mu_{p-k}\alpha_1\dots\alpha_k}=
 a_k\Pi^{\mu_1}\dots \Pi^{\mu_{p-k}}
\Pi^{\alpha_1}\dots\Pi^{\alpha_k}\quad ,\label{azca-lambdas}
\end{equation}
$ \Pi_{\rho_k i}$=$(\phi^*(\Pi_{\rho_k})_i$ and 
$\Lambda^{\rho_k i}$ corresponds to
$a_k \epsilon^{ij_1 \dots j_p}
   \Pi^{\mu_1}_{j_1} \dots \Pi^{\mu_{p-k}}_{j_{p-k}}
      \Pi^{\alpha_1}_{j_{p-k+1}} \dots \Pi^{\alpha_k}_{j_p}$
(the constants $a_k$ are fixed by $d{\tilde b}=h$).

Given the group law $g'' = g'g$ ($g''^A = g''^A (g', g)$, $A=(\alpha,
\mu;\rho_k)\,$) of $\tilde{\Sigma}$, the LI one-forms $\Pi^{A} (g)$ and the
RI vector fields $Z_{A} (g)$ are given by
\begin{equation}
\Pi^{A} (g) = \Pi^{A}_B (g) \, dg^B =  \left. \frac{\partial g''^{A}
(g',g)}{\partial g^B} \right|_{g' = g^{-1}} dg^B \; , \quad
Z_A(g) = \left. 
\frac{\partial g''^D (g',g)}{\partial g'^A} \right|_{g' = e}
\frac{\partial}{\partial g^D} 
\label{azca-LIFORM}
\end{equation}
(see, {\it e.g.}, \cite{azca-AI95}). The $Z_A$ generate the left 
$g^A$-translations of ${\tilde{\Sigma}}$. 

The {\it ${\tilde{\cal L}}_{WZ}(\xi)$ 
contribution to $j_A^i(\xi)=j_{A (kin)}^i(\xi)+j_{A (WZ)}^i(\xi)$} is
\begin{equation}
j_{A (WZ)}^i(\xi) = (\delta_A g^B)
\frac{\partial{\tilde{\cal L}}_{WZ}}{\partial g^B,_i}\equiv
(Z_A . g^B) \frac{\partial \tilde{\cal L}_{WZ}}{\partial
g^B,_i } \quad .\label{azca-tnoether}
\end{equation}
Let the extended superspace index refer to a new coordinate,
$A=\sigma_l$, and let us compute $j^i_{\sigma_l}$. 
Since only ${\tilde {\cal L}}_{WZ}$
depends on the new coordinates, $j^i_{\sigma_l (kin)}=0$. The $B$ 
summation in (\ref{azca-tnoether}) is reduced to a summation over
the additional coordinates index $\eta_k$ since the vector fields 
$Z_{\sigma_l}$ do not have $\partial/\partial x^\mu$, 
$\partial/\partial \theta^\alpha$ components and thus $Z_{\sigma_l}.g^B=0$ 
for $g^B = (\theta^\alpha,x^\mu)$. Moreover, since 
$\Lambda^{\rho_k}= \Lambda^{\rho_k}(\Pi^\mu,\Pi^\alpha)$ and 
$\Pi^\mu,\Pi^\alpha$ are defined on the standard $\Sigma$, the 
$\Lambda^{\rho_k i}$ part does not depend on $\varphi_{\eta_k}$
($g^{\eta_k}$ in (\ref{azca-tnoether})),
\begin{equation}
j_{\sigma_l}^i = (Z_{\sigma_l} . g^{\eta_k})\left( \frac{\partial}{\partial
g^{\eta_k},_i} \Pi_{\rho_k i}\right) \Lambda^{\rho_k i}\quad .
\label{azca-newcurr1}
\end{equation}
Using Eq. (\ref{azca-LIFORM}),
\begin{equation}
j_{\sigma_l}^i = (Z_{\sigma_l} . g^{\eta_k}) \left\{
\frac{\partial}{\partial g^{\eta_k},_i} \left( \left.
\frac{\partial g''_{\rho_k} (g', g)}{\partial g^B} \right|_{g'= g^{-1}}
g^B,_i \right) \right\} \Lambda^{\rho_k i} 
\left.= (Z_{\sigma_l} . g^{\eta_k}) \frac{\partial g''_{\rho_k}
 (g', g)}{\partial g^{\eta_k}} \right|_{g' = g^{-1}}
\Lambda^{\rho_k i} 
\label{azca-newcurr2}
\end{equation}
since $g{''}\neq g{''}(g,_i)$. This gives the
{\it general expression for the Noether currents associated
with the additional generators}:
\begin{equation}
j^i_{\sigma_l} = ( Z_{\sigma_l} . g''_{\rho_k} (g',g) |_{g' = g^{-1}} )
\Lambda^{\rho_k i} \equiv T_{\sigma_l \rho_k} \Lambda^{\rho_k i} \quad ,
\label{azca-adjcurr}
\end{equation}
where $T$ corresponds to the adjoint representation $Ad(g^{-1})$ and 
depends on $\xi$ through $g(\xi)$ (notice that if $X^R$ is RI 
and $\Pi^L$ as a LI one-form, $i_{X^R}\Pi^L=Ad(g^{-1}) X^R$).
Since for $A=\sigma_l$ we may restrict $D$ to $\eta_k$
($Z^{\mu,\alpha}_{\sigma_l}(g)=0)$, 
eq. (\ref{azca-adjcurr}) may also be written as
\begin{equation}
j_{\sigma_l}^i(\xi) =  \left( \left. \frac{\partial g''^{\eta_k} (g', g)}
{\partial g'^{\sigma_l}} \right|_{g' = e}  
\left. \frac{\partial g''_{\rho_k} (g', g)}{\partial g^{\eta_k}} 
\right|_{g' = g^{-1}} \right) \Lambda^{\rho_k i} 
\label{azca-curr2}
\end{equation}
using (\ref{azca-LIFORM}); the bracketed term is determined by the 
group ${\tilde \Sigma}$ only, and $\Lambda^{\rho_k i}$ by 
${\tilde {\cal L}}_{WZ}(\xi)$.

\noindent a) {\it D=10, N=1 superstring}:

 Using expression (\ref{azca-ssWZt}) for (\ref{azca-WZinvar}), we 
find that the conserved Noether currents are 
\begin{equation}
     j^{\mu i}_{(\varphi)}=\epsilon^{ij}\partial_j x^\mu
  \quad ,\quad j^{\alpha i}=(1/2)\epsilon^{ij}\partial_j\theta^\alpha \ ,
\end{equation}
and the charges \cite{azca-AGIT89}
\begin{equation}
Z^\mu = \oint d\sigma j^{\mu 0}_{(\varphi)}=
\oint d \sigma \, {\partial x^\mu \over \partial \sigma}\ ,
\quad
Z^\alpha = \oint d\sigma j^{\alpha 0}= 
  \oint d\sigma \frac{1}{2}\frac{\partial\theta^\alpha}{\partial\sigma}
  =0 \ ,
  \label{azca-sscc}
\end{equation}
assuming that $\theta$ is periodic
in $\sigma$ ({\it cf}. \cite{azca-MHMS00}). It is clear that, in general, the 
integral of $j^0$ (as, {\it e.g.}, for  $j^{\mu 0}_{(\varphi)}$)
leads to a non-zero result if the topology is nontrivial (the loop
is not contractible).

\noindent b) {\it D=11 2-brane}:

It can be shown from eq. (\ref{azca-smbt}) that the currents can be 
written as the worldvolume duals of the {\it current two-forms}
\begin{eqnarray}
J^{\mu \nu} 
 &=& d \left( {2 \over 3} x^{[\mu} dx^{\nu]} 
       + {1 \over 15} \theta^\alpha x^{[\mu} 
         (C \Gamma^{\nu]})_{\alpha \beta} d\theta^\beta \right)
\quad,
\nonumber \\
J^{\kappa \alpha} 
 &=&  d \left({3 \over 5} dx^{\kappa}  \theta^{\alpha}
       - {1 \over 30} (C \Gamma^{\kappa})_{\beta \gamma}
       \theta^{\beta}  \theta^{\alpha} d
       \theta^{\gamma} \right)
\, ,
\quad
J^{\beta \gamma} =  d (-{2 \over 15}\theta^{\beta}
                            d\theta^{\gamma})
\quad ;
\label{azca-JAsm}
\end{eqnarray}
current conservation follows from $d J=0$. For periodic $\theta$'s 
the charges $Z^{\kappa_1 \alpha_1}$, $Z^{\beta_1 \gamma_1}$
turn out to be zero, but not
$Z^{\mu_1\nu_1}$ for a non-trivial closed two-cycle \cite{azca-AGIT89}
(in the general $p$-case, the integrals
are over non-trivial de Rham $p$-cycles; we refer to \cite{azca-AGIT89}
for details on topological charges). Thus, the above assumptions 
provide a realization of the extended algebra where
only the bosonic $Z^{\mu\nu}$ generator is realized non-trivially.

%%%%%%%%%%%%%%%%%%%%%%%%%%%%%%%%%%%%%%%%%%%%%%%%%%%%%%%%%%%%%%%%
%%%%%%%%%%%%%%%%%%%%%%%%%%%%%%%%%%%%%%%%%%%%%%%%%%%%%%%%%%%%%%%%
\section{The case of D-branes}
\label{azca-casedbr}

Consider first a bosonic background such that the action 
of the D$p$-brane \cite{azca-P95,azca-Dp} reduces to
\begin{equation}
 I= \int d^{p+1} \xi \sqrt{-\det(\partial_i x^\mu \partial_j x_\mu
+F_{ij})}
\quad ,
\label{azca-borninfeld}                                    
\end{equation} 
where $F=dA$ and $A(\xi)=A_i(\xi)d\xi^i$ is the worldvolume Born-Infeld (BI)
field. 

Let us look for a {\it manifestly supersymmetric generalisation}.
This means substituting first $\Pi^\mu_i$ for $\partial_ix^\mu$, 
$F_{ij}=\partial_{[i}A_{j]}$ by
${\cal F}=dA-B$, and then adding a WZ term $b$, $db=h$.
A previous analysis \cite{azca-AT89} of the WZ terms of the scalar 
branescan \cite{azca-AETW87} showed that {\it WZ terms may be characterized 
and classified by CE-(p+2)cocycles}. The same philosophy is successful
for the D$p$-branes. The result is that D$p$-branes may also be be 
characterized (see below and \cite{azca-CAIP00} for details and 
further references) by means of non-trivial CE $(p+2)$-cocycles, 
recovering Polchinski's consistency conditions \cite{azca-P95} 
($p$ even/odd for IIA/IIB). In the case of D-branes, however,
and due to the presence of $F_{ij}$ in the kinetic term 
(\ref{azca-borninfeld}) the situation turns out to be 
different from that of the previous $p$-branes: the new variables 
will appear in the action {\it non-trivially}, not as total 
derivatives.

%%%%%%%%%%%%%%%%%%%%%%%%%%%%%%%%%%%%%%%%%%%%%%%%%%%%%%%%%%%%%%%%%%%%%%%%%%%%%%%%

\subsection{Example: the D2-brane defined on its extended superspace}
\label{azca-Dbext}

Consider the D2-brane. The starting point is now the IIA-type FDA 
plus the  $d{\cal F}$ equation {\it i.e.}
\be
\addtolength{\arraycolsep}{-.5ex}
\begin{array}{rclcrcl}
\dis
d\Pi^\alpha&=&0
& \qquad \quad &
\dis
d\Pi^\mu &=& \frac{1}{2}(C\Gamma^\mu)_{\alpha\beta}\Pi^\alpha\Pi^\beta
\\[1.3ex]
\dis
d\Pi &=& \frac{1}{2}(C\Gamma_{11})_{\alpha\beta}\Pi^\alpha\Pi^\beta
& \qquad \quad &
\dis
d\Pi_{\mu\nu} &=& \frac{1}{2}(C\Gamma_{\mu\nu})_{\alpha\beta}\Pi^\alpha
                  \Pi^\beta
\\[1.3ex]
\dis
d\Pi_\mu^{(z)} &=& \frac{1}{2}(C\Gamma_\mu\Gamma_{11})_{\alpha\beta}
                   \Pi^\alpha\Pi^\beta 
& \qquad \quad &
\dis
d{\cal F}&=& (C\Gamma_\mu\Gamma_{11})_{\alpha\beta}\Pi^\mu\Pi^\alpha
              \Pi^\beta   \, ,
 \label{azca-freedbrane}
\end{array}
\ee
$(\mu= 0,\dots 9$, $\alpha=1,\dots 32$). This is justified {\it e.g}
by the fact that the dual of the first $5$ eqs. is the algebra 
obtained when one computes the algebra of Noether charges
for the type IIA D$2$-brane \cite{azca-H98}. The next step is 
extending this algebra with the generators obtained by replacing 
vector indices by spinorial ones, as outlined after (\ref{azca-dPim1a1}). 
In the case of the D2-brane this is not difficult 
to do because, apart from the equation for $d{\cal F}$, the FDA above 
is actually the dimensional reduction to D=10 of the D=11 one
(eq. (\ref{azca-2.10}) with generators with one or two vector indices
since $p$=2),
\begin{equation}
d\Pi^{\tilde\mu}
   =(1/2)(C\Gamma^{\tilde\mu})_{\alpha\beta}\Pi^\alpha\Pi^\beta
\, ,\qquad \quad
d\Pi_{\tilde\mu\tilde\nu}
   =(1/2)(C\Gamma_{\tilde\mu\tilde\nu})_{\alpha\beta}\Pi^\alpha
  \Pi^\beta  \, ,
\label{azca-freeeleven}
\end{equation}
where (${\tilde\mu}=(\mu,10)=0,1,\dots 10$), and in which 
one sets $\Pi^{\tilde\mu}\equiv(\Pi^\mu,\Pi^{10}\equiv\Pi)$,
$\Pi_{\tilde\mu\tilde\nu}
\equiv(\Pi_{\mu\nu},\Pi_{\mu 10}\equiv\Pi_\mu^{(z)})$. 
This $D=11$ FDA may be extended. The $D=10$ dimensional reduction 
of the extended algebra gives
\begin{eqnarray}
   d\Pi^\alpha &=& 0\ ,
    \quad
     d\Pi^\mu=\frac{1}{2}(C\Gamma^\mu)_{\alpha\beta}\Pi^\alpha\Pi^\beta\ ,
   \quad
     d\Pi=\frac{1}{2}(C\Gamma_{11})_{\alpha\beta}\Pi^\alpha\Pi^\beta\ ,
    \nonumber\\
     d\Pi_{\mu\nu} &=& \frac{1}{2}(C\Gamma_{\mu\nu})_{\alpha\beta}\Pi^\alpha
    \Pi^\beta\ ,
     \quad
     d\Pi_\mu^{(z)}=\frac{1}{2}(C\Gamma_\mu\Gamma_{11})_{\alpha\beta}
       \Pi^\alpha\Pi^\beta \ ,
      \nonumber \\
   d\Pi_{\mu\alpha}&=&(C\Gamma_{\nu\mu})_{\alpha\beta}\Pi^\nu\Pi^\beta+
    (C\Gamma_{11}\Gamma_\mu)_{\alpha\beta}\Pi\Pi^\beta+
     (C\Gamma^\nu)_{\alpha\beta}\Pi_{\nu\mu}\Pi^\beta-(C\Gamma_{11})_{\alpha\beta}
    \Pi_\mu^{(z)}\Pi^\beta\quad ,
    \nonumber
   \\
  d\Pi_\alpha^{(z)}&=&(C\Gamma_\nu\Gamma_{11})_{\alpha\beta}\Pi^\nu\Pi^\beta
   +(C\Gamma^\nu)_{\alpha\beta}\Pi_\nu^{(z)}\Pi^\beta\quad ,
    \nonumber
    \\
    d\Pi_{\alpha\beta}&=&-\frac{1}{2}(C\Gamma_{\mu\nu})_{\alpha\beta}
     \Pi^\mu\Pi^\nu-
   (C\Gamma_\mu\Gamma_{11})_{\alpha\beta}\Pi^\mu\Pi-\frac{1}{2}
   (C\Gamma^\mu)_{\alpha\beta} \Pi_{\mu\nu}\Pi^\nu
   \nonumber
\\
    & &   +\frac{1}{2}(C\Gamma_{11})_{\alpha\beta}
   \Pi_{\mu}^{(z)}\Pi^\mu
    -\frac{1}{2}(C\Gamma^\mu)_{\alpha\beta}
     \Pi_\mu^{(z)}\Pi
   +\frac{1}{4}(C\Gamma^\mu)_{\alpha\beta}\Pi_{\mu\delta}\Pi^\delta
\nonumber
\\
 & & +   \frac{1}{4}(C\Gamma_{11})_{\alpha\beta}\Pi_\delta^{(z)}\Pi^\delta
   +2\Pi_{\mu(\beta}(C\Gamma^\mu)_{\alpha)\delta}\Pi^\delta+
    2(C\Gamma_{11})_{\delta(\alpha}\Pi_{\beta)}^{(z)}\Pi^\delta
  \quad .
                                    \label{azca-freedbraneext}
\end{eqnarray}
Using the new forms it is possible to find a manifestly invariant
WZ form ${\tilde b}$, $d{\tilde b}=h$; $h$ is given by
\begin{equation}
      h=(C\Gamma_{\mu\nu})_{\alpha\beta}\Pi^\mu\Pi^\nu\Pi^\alpha\Pi^\beta
    -(C\Gamma_{11})_{\alpha\beta}\Pi^\alpha\Pi^\beta {\cal F}
                         \quad ,             \label{azca-hagain}
\end{equation}
and the {\it manifestly invariant WZ term for the type IIA D2-brane} by
\begin{equation}
  {\tilde b} =
 \frac{2}{3}\Pi_{\mu\nu}\Pi^\mu\Pi^\nu+\frac{4}{3}\Pi_\mu^{(z)}\Pi^\mu
    \Pi-\frac{2}{15}\Pi_{\alpha\beta}\Pi^\alpha\Pi^\beta
   -\frac{3}{5}\Pi_{\mu\alpha}\Pi^\mu\Pi^\alpha-\frac{3}{5}\Pi_\alpha^{(z)}
   \Pi\Pi^\alpha-2\Pi {\cal F}\ .
                                 \label{azca-bdbrane}
\end{equation}
We expect that this analysis also holds true for the other values of $p$.

The extended
free differential algebra is not the dual of a Lie algebra 
because it includes the equation for the {\it three}-form $d{\cal F}$.
However, 
\begin{equation}
      d(\frac{1}{2}\Pi^\alpha\Pi_\alpha^{(z)}-\Pi^\mu\Pi_\mu^{(z)})=
    (C\Gamma_\mu\Gamma_{11})_{\alpha\beta}\Pi^\mu\Pi^\alpha
      \Pi^\beta                                \label{azca-6.22}
\end{equation}
so that, on the extended superspace ${\tilde \Sigma}(\theta^\alpha,x^\mu,
\varphi_\mu,\varphi_\alpha)$ we may set
\begin{equation}
{\cal F}=
(1/2)\Pi^\alpha\Pi_\alpha^{(z)}-\Pi^\mu\Pi_\mu^{(z)}\quad.
\label{azca-fonsuper}
\end{equation}
 Since ${\cal F}=dA-B$ {\it and $B$ is defined on $\Sigma$}, it
follows that $dA$ may be written on ${\tilde \Sigma}$. 
Making use of the explicit form of the LI one-forms in terms of the extended 
superspace variables,
it is easy to identify $A$ {\it as the one-form on} $\tilde\Sigma$
\begin{equation}
    A=\varphi_\mu dx^\mu+(1/2)\varphi_\alpha d\theta^\alpha \quad .
                                                  \label{azca-newA}
\end{equation}
In the present approach,
the customary BI worldvolume field $A_i(\xi)d\xi^i$ 
becomes $\phi^*(A)$; we might even say that the existence of the BI 
field is a consequence of supersymmetry. We now check the consistency 
of the replacement (\ref{azca-newA}).

 a) The  Euler Lagrange equations are still
the same.
Let $I[x^\mu(\xi),\theta^\alpha(\xi),A_i(\xi)]$ be the action before making the 
substitution. The EL equations are
\begin{equation}
     \delta I/\delta x^\mu=0
\, ,\qquad 
\delta I/\delta \theta^\alpha=0
\, ,\qquad
\delta I/\delta A_j=0 \, .
\label{azca-ELbefore}
\end{equation}  
When the substitution is made, 
\begin{equation}
\label{azca-elr}
 \begin{array}{rclcrcll}
\displaystyle
\int d\xi{'}^{p+1}\frac{\delta I}{\delta A_j(\xi{'})}
\frac{\delta A_j(\xi{'})}{\delta x^\mu(\xi)}+\frac{\delta I}{\delta 
x^\mu} & = & 0
\, , & \; &
\displaystyle
\frac{\delta I}{\delta\varphi_\mu} =
\frac{\delta I}{\delta A_j}\partial_j x^\mu & = & 0
\\[2.6ex]
\displaystyle
\int d\xi{'}^{p+1}\frac{\delta I}{\delta A_j(\xi{'})}
\frac{\delta A_j(\xi{'})}{\delta \theta^\alpha(\xi)}+
\frac{\delta I}{\delta \theta^\alpha} &= & 0
\, , & \;  &
\displaystyle
   \frac{\delta I}{\delta\varphi_\alpha}=
     \frac{1}{2} \frac{\delta I}{\delta A_j}\partial_j
\theta^\alpha & = & 0 \, .
  \end{array}\label{azca-8.22}
\end{equation}
We see that to avoid the collapse of one or more worldvolume 
dimensions we must have $\delta I/\delta A_j=0$ which
implies eqs. (\ref{azca-ELbefore}). This also follows from the
fact that $\delta I/\delta \varphi_\mu=0$ implies 
$(\delta I/\delta A_j)g_{ij}=0$, where  $g_{ij}\equiv\Pi^\mu_i\Pi_{\mu j}=
\partial_i x^\mu\partial_j x_\mu+\textrm{(nilpotent terms)}$ is 
the induced worldvolume metric. Thus, we must have $\delta I/\delta A_j=0$
to prevent $g_{ij}$ from being degenerate. As a result, 
$\delta I/\delta\varphi_\alpha=0$ is satisfied identically and
it is a Noether identity.

b) The gauge transformations of $A_i(\xi)$ can be reinterpreted 
in the new language. If one defines
$\delta \varphi_\mu= \partial_\mu \lambda$ and $\delta \varphi_\alpha=
2\partial_\alpha \lambda$, by means of a superfield $\lambda$ such that
$\phi^* \lambda(x^\mu,\theta^\alpha)=\Lambda(\xi)$, then $\phi^*(A)$ 
behaves as expected: $\delta(\phi^* [\varphi_\mu dx^\mu+
\frac{1}{2}\varphi_\alpha d\theta^\alpha])=\partial_i \Lambda$.

 c) The number of worldvolume degrees of freedom remains the same.
Let us first note that, since $\delta I/\delta\varphi_\alpha=0$ 
is a Noether identity, the second Noether theorem 
tells us that there exists a gauge symmetry that can be used 
to set $\varphi_\alpha=0$. Thus, the `physical' part of $A$ is 
contained in $\varphi_\mu dx^\mu$. 
The identification (\ref{azca-newA}) is therefore equivalent to 
replacing $A_i(\xi)$ by $\varphi_\mu(\xi)\partial_i x^\mu(\xi)$. 
We now notice that the $D$ equations $\delta I/\delta \varphi_\mu=0$
produce only $(p+1)$ independent ones, $\delta I/\delta A_j=0$;
the remaining $D-(p+1)$ equations are Noether identities that reflect
the existence of further gauge symmetries. To check explicitly
the degrees of freedom we first adopt the gauge
$(x^0(\xi)=\tau,\, x^1(\xi)=\xi^1,...,x^p(\xi)=\xi^p)$. Then,
\begin{equation}
    (\phi^*A)_i = \varphi_\mu(\xi)\partial_i
x^\mu(\xi)=\varphi_i(\xi)+\varphi_{K}(\xi)\partial_i x^{K}(\xi)
\ ,\quad K=p+1,\dots,D-1 \ .\label{azca-dof1}
\end{equation}
We see that, apparently, we are describing the $(p+1)$ components
$A_i$ of the BI field using $D$ functions $(\varphi_i,\varphi_{K})$. 
The mismatch in the number of degrees of freedom is sorted out by the 
existence of the bosonic gauge symmetries that allow us to remove 
the additional ($D-p-1$) functions. Futhermore, since the components 
$\varphi_\mu$ enter 
the action non-trivially only through $(\phi^*A)_i$, any local 
transformation of $\varphi_\mu(\xi)$ that leaves  $(\phi^*A)_i$
unchanged will be a gauge symmetry of the action. Consider then
\begin{equation}
  \delta\varphi_i(\xi)=-\alpha_K(\xi)\partial_i x^{K}(\xi)\ ,\quad
   \delta\varphi_K(\xi)=\alpha_k(\xi)\ .
                                   \label{azca-dof2}
\end{equation}
This specific transformation has the property $\delta(\phi^*A)_i=0$, so it 
is a gauge symmetry that can be used to set $\varphi_{K}=0$ by taking
$\alpha_k=\varphi_K$, so $\phi^*(A)_i=
\varphi_i$. Hence, we may identify $A_i=\phi^*(A)_i$.

For the 
IIB D$p$-brane, an analysis similar to that in this
section (for odd $p$) can be made. In fact, the origin of $A(\xi)$ 
in the $p$=1 IIB D-string case was discussed in
\cite{azca-S99} (see also \cite{azca-S00}) 
by introducing an appropriate extended group manifold.
We may conclude, then, that 
{\it the different worldvolume fields
are introduced naturally through the pull-back of coordinates (forms)
of (defined on) suitably extended superspaces.}

\section{Noether charges and D-brane actions}
\label{azca-2bn}

The worldvolume field $A(\xi)$ that appears in the D2-brane action 
may be written in terms of the variables of the superstring extended 
superspace ${\tilde \Sigma}(x^\mu,\theta^\alpha,\varphi_\mu, \varphi_\alpha)$.
The D2-WZ term, which is quasi-invariant in 
these coordinates, can be made strictly invariant by further extending the 
previous superspace to
${\tilde \Sigma}=(x^\mu,\theta^\alpha,\varphi_\mu,
\varphi_\alpha$, $\varphi_{\mu\nu},\varphi_{\mu\alpha},\varphi_{\alpha\beta},
\varphi)$. In this way, the whole action is invariant.
The canonical commutators of the charges generating the symmetries of
the action (denoted by a hat) give a realization of the `right' version of 
the `left' Lie algebra dual to (\ref{azca-freedbraneext}). 

Consider the $\{ Q_\alpha, Q_\beta\}$ commutator, that we shall write as
\begin{equation}
 \{ Q_\alpha,Q_\beta\}= (C\Gamma^\mu)_{\alpha\beta}P_\mu +
  (C\Gamma_\mu\Gamma_{11})_{\alpha\beta}{\hat Z}^\mu +
  (C\Gamma_{\mu\nu})_{\alpha\beta}{\hat Z}^{\mu\nu}
  + (C\Gamma_{11})_{\alpha\beta}{\hat Z}\quad   .
  \label{azca-QQ}
\end{equation} 
With $A=A(\xi)$,
the $C\Gamma_{\mu\nu}$ and $C\Gamma_{11}$ contributions
would come from the quasi-invariance of the WZ Lagrangian, while
$C\Gamma_{11}\Gamma_\mu$ would be the result of the contribution of the
$A(\xi)$ field to the Noether current \cite{azca-H98} 
(see also \cite{azca-BT98}). This is because the supersymmetry 
transformations {\it do not close on $A$}, and this produces an 
additional term by a mechanism similar to the one in the standard 
quasi-invariance case.

These modifications become transparent
by formulating the action on the extended superspace \cite{azca-CAIP00}.    
Consider the formulation of the D2-brane on the extended superspace 
with quasi-invariant WZ term
$b=b(x^\mu,\theta^\alpha,\varphi_\mu,\varphi_\alpha)$. The conserved 
Noether currents then {\it have to} include a term coming from
the quasi-invariance of the WZ
piece: if we wrongly ignored this  the 
 algebra of the charges would be 
\begin{equation}
\{ Q_\alpha,Q_\beta\}= (C\Gamma^\mu)_{\alpha\beta}P_\mu +
  (C\Gamma_\mu\Gamma_{11})_{\alpha\beta}{\hat Z}^\mu   
\label{azca-2Db}
\end{equation} 
rather than (\ref{azca-QQ}). 
Alternatively, we may find the correct algebra 
by replacing the quasi-invariant WZ term $b$ by 
${\tilde b}={\tilde b}(x^\mu,\theta^\alpha,\varphi_\mu,\varphi_\alpha,
\varphi_{\mu\nu},\varphi_{\mu\alpha},\varphi_{\alpha\beta},\varphi)$,
which is {\it manifestly invariant} since the transformation properties 
of the additional variables $(\varphi_{\mu\nu},
\varphi_{\mu\alpha},\varphi_{\alpha\beta},\varphi)$ remove  the 
quasi-invariance of the WZ term $b$. 
Hence,  the algebra of charges reproduces (\ref{azca-QQ}), and 
the contributions to ${\hat Z}^{\mu\nu}$ and $\hat Z$ are 
entirely due to the contribution of the additional variables
$\varphi_{\mu\nu}$, $\varphi_{\mu\alpha}$, $\varphi_{\alpha\beta}$,
$\varphi$ in the WZ term $\tilde b$ (or to the quasi-invariance 
of $b(x^\mu,\theta^\alpha,\varphi_\mu,\varphi_\alpha)$ if we
used $b$ instead). 

%%%%%%%%%%%%%%%%%%%%%%%%%%%%%%%%%%%%%%%%%%%%%%%%%%%%%%%%%%%%%%%%%%%%%%%
%%%%%%%%%%%%%%%%%%%%%%%%%%%%%%%%%%%%%%%%%%%%%%%%%%%%%%%%%%%%%%%%%%%%%%%
%%%%%%%%%%%%%%%%%%%%%%%%%%%%%%%%%%%%%%%%%%%%%%%%%%%%%%%%%%%%%%%%%%%%%%%
\section{Higher order tensors: the case of the M5-brane}
\label{azca-m5b}
%%%%%%%%%%%%%%%%%%%%%%%%%%%%%%%%%%%%%%%%%%%%%%%%%%%%%%%%%%%%%%%%%%%%%%%

Consider the $D=11$ M5-brane, which contains a 
worldvolume two-form field $A(\xi)$. As before, the 
supersymmetric action is obtained in two steps:
 
a) First, 
$H=dA-C$ where $C$ is such that
$
    dC=-(C\Gamma_{\mu\nu})_{\alpha\beta}\Pi^\mu\Pi^\nu\Pi^\alpha\Pi^\beta
 $,
and the transformation properties of $A$ are fixed so 
that $H$ is invariant; 

b) Secondly, a WZ term is added to
obtain $\kappa$-symmetry.

The FDA generated by the LI 
one-forms $\Pi^\alpha$,
$\Pi^\mu$ and the three-form $H$ is
\begin{equation}
d\Pi^\alpha=0
\quad ,
\quad
d\Pi^\mu=(1/2)(C\Gamma^\mu)_{\alpha\beta}\Pi^\alpha\Pi^\beta\quad ,
\quad
dH=(C\Gamma_{\mu\nu})_{\alpha\beta}\Pi^\mu\Pi^\nu\Pi^\alpha\Pi^\beta
\, .
\label{azca-11b}
\end{equation}
(Note that $ddH\equiv 0$ implies $(C\Gamma^{\mu\nu})_{(\alpha\beta}
(C\Gamma_\nu)_{\gamma\delta)}\equiv 0$ which is satisfied for $D=11$).
To find the nontrivial CE $(p+2)$-cocycles for the FDA (\ref{azca-11b}) 
one may impose the closure condition for $h$  on a general $(p+2)$-form 
with the correct dimensions. This gives two possible 
expressions for $h$. One of them is proportional to
$(C\Gamma_{\mu\nu})_{\alpha\beta}\Pi^{\mu}\Pi^{\nu}\Pi^\alpha\Pi^\beta=dH$, 
so it is exact. The other is found to be
\begin{equation}
      h\propto (C\Gamma_{\mu_1\dots\mu_5})_{\alpha\beta}
         \Pi^{\mu_1}\dots\Pi^{\mu_5}\Pi^\alpha\Pi^\beta -(15/2)
         (C\Gamma_{\mu_1\mu_2})_{\alpha\beta}
         \Pi^{\mu_1}\Pi^{\mu_2}\Pi^\alpha\Pi^\beta H  \ ,      \label{azca-11g}
\end{equation}
which turns out to be not CE-exact. Hence, there is no solution 
unless $p=5$: {\it the M5-brane, $p=5$, is characterized by the only 
non-trivial D=11 ($5$+$2$)-CE-cocycle}.

$H$ may be defined as a LI three-form on the extended superspace group
of coordinates\break 
${\tilde\Sigma}(\theta^\alpha,x^\mu,\varphi_{\mu\nu},\varphi_{\mu\alpha},
\varphi_{\alpha\beta})$, namely
\begin{equation}
     H=\frac{2}{3}\Pi^\mu\Pi^\nu\Pi_{\mu\nu}+\frac{3}{5}\Pi^\mu\Pi^\alpha
   \Pi_{\mu\alpha}-
      \frac{2}{15}\Pi_{\alpha\beta}\Pi^\alpha \Pi^\beta      
                 \quad .                                         \label{azca-11i}  
\end{equation}
Moreover, it may be shown that there exists a LI $\tilde b$ such that 
$h=d{\tilde b}$ on a suitably extended superspace \cite{azca-S97}.
By using the explicit form of the LI one-forms appearing in
(\ref{azca-11i}),
we may replace the worldvolume two-form $A(\xi)$ 
by a two-form $A$ on the extended 
superspace. Also, the gauge transformation 
$\delta A(\xi)=d\Lambda(\xi)$ is achieved by the 
one-form  $\lambda=\lambda_\mu dx^\mu+\lambda_\alpha d\theta^\alpha$, 
$\phi^*(\lambda)=\Lambda(\xi)$. Then, defining $\delta\varphi_{\mu\nu}$,
$\delta\varphi_{\mu\alpha}$, $\delta\varphi_{\alpha\beta}$ conveniently
one obtains $\delta\phi^*(A)=d\Lambda(\xi)$.

The EL equations derived from $I[x^\mu(\xi),\theta^\alpha(\xi),A_{ij}(\xi)]$
are equivalent to the ones corresponding to the new action in which 
$A(\xi)$ is the pull-back $\phi^*(A)$. In fact, in parallel with the 
D2 brane case, it is found that ${\delta I}/{\delta\varphi_{\alpha\beta}}=0$ 
and ${\delta I}/{\delta\varphi_{\mu\alpha}}=0$ are identically 
satisfied (they are Noether identities) and that only
${\delta I}/{\delta\varphi_{\mu\nu}}=0$ contains a non-trivial
part, $\delta I/\delta A_{ij}=0$. Thus, there remain
${D\choose 2} - {p+1\choose 2}$ Noether identities. Consider then 
$(\phi^*(\varphi_{\mu\nu} dx^\mu dx^\nu))_{ij}$ = 
$A_{ij}(\xi)=\varphi_{\mu\nu}\partial_i x^\mu\partial_j x^\nu$.
Again, the election $x^0=\tau$, $x^1(\xi)=\xi^1,\dots,x^5(\xi)=\xi^5$,
gives
\begin{equation}
   (\phi^*A)_{ij}=\varphi_{\mu\nu}
\partial_i x^\mu\partial_j x^\nu=\varphi_{ij}(\xi)+\varphi_{iK}(\xi)
\partial_j x^K-\varphi_{jK}\partial_i x^K+
\varphi_{KL}\partial_i x^K\partial_j x^L \quad ,
\label{azca-m5dof1}
\end{equation}
where $K,L=((p+1)=6,\dots,D-1=10$). The additional degrees of freedom 
associated with $\varphi_{iK}$ and $\varphi_{KL}$ may be removed by 
suitable gauge transformations. Indeed,
\begin{eqnarray}
     \delta_\alpha\varphi_{ij} &=& 0\quad ,\quad \delta_\alpha
\varphi_{iK}=\frac{1}{2}\alpha_{KL}\partial_i x^L\quad ,\quad
\delta_\alpha\varphi_{KL}=\alpha_{KL}\quad , \nonumber\\
\delta_\beta\varphi_{ij} &=& -\beta_{iK}\partial_j x^K
+\beta_{jK}\partial_i x^K\quad ,\quad \delta_\beta
\varphi_{iK}=\beta_{iK}\quad ,\quad
\delta_\beta\varphi_{KL}=0\quad ,   
\label{azca-m5dof2}
\end{eqnarray}
leave $(\phi^*A)_{ij}$ invariant, $\delta_\alpha$ 
removes $\varphi_{KL}$ (by choosing $\alpha_{KL}=-\varphi_{KL}$), 
and $\delta_\beta$ sets $\varphi_{iK}$ equal to zero 
(for $\beta_{iK}= -\varphi_{iK}$).

The previous discussions of the degrees of freedom for the  D2 and M5 
worldvolume fields set the pattern for other possible cases.

\section{Conclusions}

In view of the results described here, it seems natural to 
conclude that there exists an extended superspace origin for all 
the worldvolume fields appearing in the various super-$p$-brane 
actions: all worldvolume fields may be considered as pull-backs to 
$W$ for the map $ \phi:W\longrightarrow {\tilde\Sigma}$. In other words,
{\it there exists a field/extended superspaces democracy} by which
{\it all superbrane worldvolume fields may be seen as the 
pullbacks $\phi^*$ to $W$ of some target extended superspace $\tilde\Sigma$
coordinates}.

The appropriate extended superspace 
${\tilde \Sigma}$ of the specific theory being considered
is determined by an extension of its associated {\it basic} sTr${}_D$
{\it fermionic group} and, using ${\tilde \Sigma}$, the action of the
super-$p$-brane can be 
constructed in a manifestly invariant form. In fact, 
{\it in this field/extended 
superspace democracy context, 
 the invariance properties 
and the non-trivial cocycles of the CE cohomology 
 appear to characterise essentially 
the different superbranes and their actions}
\cite{azca-CAIP00}
(we might also say that they are {\it perfect} in the sense
of \cite{azca-GRS81}).

Are these extra `dimensions' {\it necessary} or just {\it convenient} for
a more geometrical and unified description of superbranes? We already
saw that spacetime itself $(x^\mu)$ {\it is}
a {\it consequence} of the {\it non-triviality} of the D-Minkowski 
space-valued second cohomology group of the abelian odd 
translation group $\textrm{sTr}_D$. Thus, it is reasonable to
conclude that supersymmetry algebras and superspace groups going 
beyond the standard ones (see {\it e.g.} 
\cite{azca-AGIT89,azca-G89,azca-BS95,azca-S99,azca-CUR88,azca-AIT91,
azca-B96, azca-S97,azca-DG97,azca-B97}) are {\it required} for a suitable 
description of the various superstring and superbrane theories and that, 
as in the superspace case, Nature makes use of the extension possibilities
offered by the non-trivial cohomology groups of $\textrm{sTr}_D$.

\vskip 1cm
{\bf Acnowledgements}.
This work has been partially supported by the DGICYT research grant PB 
96-0756 and the Junta de Castilla y Le\'on research grant C02/199. The 
authors wish to thank Igor Bandos and Paul Townsend
for helpful discussions.

\end{document}